\documentclass[runningheads,a4paper]{llncs}

\usepackage{amssymb}
\usepackage{graphicx}

\RequirePackage{amsmath}
\RequirePackage[hidelinks]{hyperref}
\usepackage{tikz}
\usepackage{tkz-graph}
\usetikzlibrary{arrows}
\newcommand{\bigCI}{\mathrel{\text{\scalebox{1.07}{$\perp\mkern-10mu\perp$}}}}
\usepackage{longtable}
\usepackage{booktabs}
\usepackage{array}
\usepackage[ruled]{algorithm2e}
\usetikzlibrary{decorations.markings}
\tikzstyle{vertex}=[circle, fill,draw, inner sep=0pt, minimum size=8pt]
\newcommand{\vertex}{\node[vertex]}
\begin{document}

\mainmatter  

\title{Model Selection for Graphical Log-linear Models: A Forward Model Selection Algorithm based on Mutual Conditional Independence}

\titlerunning{Model Selection for Graphical Log-linear Models}

%
%
\author{Niharika Gauraha}
\authorrunning{Niharika Gauraha}

\institute{Systems Science and Informatics Unit\\
       Indian Statistical Institute\\
       8th Mile, Mysore Road Bangalore, India\\
 niharika.gauraha@gmail.com
}

%
%

\toctitle{Lecture Notes in Computer Science}
\tocauthor{Authors' Instructions}
\maketitle

\begin{abstract}
Model selection and learning the structure of graphical models from the data sample constitutes an important field of probabilistic graphical model research, as in most of the situations the structure is unknown and has to be learnt from the given dataset. In this paper, we present a new forward model selection algorithm for graphical log-linear models. We use mutual conditional independence check to reduce the search space which also takes care of the evaluation of the joint effects and chances of missing important interactions are eliminated. We illustrate our algorithm with a real dataset example.
\end{abstract}

\section{Introduction}
Graphical models are a way of representing relationships among random variables using a graph and are adopted by a variety of research fields such as data mining, natural language processing and bioinformatics etc. The graphical models may be based on directed acyclic graphs, undirected graph and mixed graphs(see \cite{LaurtBook} for details). We mainly focus on graphical log-linear models that are undirected graphical models and are used for visual representation of log-linear models, in particular for higher dimensional tables. For further information on log-linear models we refer from \cite{BishopBook} to \cite{Andersen}.

Graphical Log Linear Models(GLLM) are a subclass of hierarchical log-linear models. Selecting an optimal model from the class of graphical models is known to be an intractable problem. Most of all existing model selection algorithms are based on forward selection, backward elimination or combination of the both. For detailed discussion on graphical log-linear model selection we refer \cite{BishopBook}, \cite{Christensen}, \cite{CorDecompModel}, \cite{Goodman3} and \cite{Wermuth}. 

Current model selection algorithms mostly use conditional independence property as model selection criteria. We propose a forward model selection algorithm which is based on the concept of equivalence between mutual conditional independence in probability and independent set in graph theory. We use mutual conditional independence check to reduce the search space for the local search algorithm.  We also use the concept that, there is a one-to-one relationship between graphs and All Maximal Independent set (AMIS). Our main focus is to find the AMIS for the underlying graphical model based on the data sample. The main drawback of the traditional forward selection algorithm is that some joint effects of the group of factors may not be evaluated thus many important interactions can be missed(see \cite{BishopBook} for details). As we perform mutual conditional independence check at every step for a group of factors that takes care of the evaluation of the joint effects and chances of missing important interactions are eliminated.

At a high level, our algorithm works as follows. Suppose we have a p-factor contingency table and a graph $G = (V,E)$ consists of p vertices $V = \{X_1,...,X_p\} $ that corresponds to the factors of the contingency table. 
Let the vertex set $\{ X_i,X_j,X_k \}$ be a maximal independent set of G. The set $\{ X_i,X_j,X_k \}$ forms an independent set if and only if the factors $\{ X_i,X_j,X_k \}$ are mutually independent conditioned on the remaining factors.  The algorithm starts with the null model(a model with complete independence) and then we use rules based on mutual conditional independence for adding edges to arrive at a smallest graphical model that fits the data. We note that here model selection is inferring the AMIS or equivalently the edge set $E$ from the data sample. We use the minimization of the deviance from the saturated model as a model selection criteria but it can be extended to many other metrics as well. 

The remainder of the paper is arranged as follows. In section 2 we briefly review the required concepts such as graph theory, conditional independence, GLLMs, model testing and model comparison criterion.  Section 3 introduces concept of mutual conditional independence. In section 4 we present new forward model selection algorithm with illustrated example. In section 5 we give computational details that we used for model selection. In Section 6 we conclude and discuss future scope. 

\section{Background and Notation}
In this section, we briefly review the required concepts.
\subsection{Graph Theory}
Here we list and define fundamental principles of graph theory that we will be using in later sections. For details on graph theory we refer to \cite{West}. 
\begin{definition}[Undirected Graphs]
A graph G, is a pair G = (V, E), where V is a set of vertices and E is a set of edges. A graph is  said to be an undirected graph if its vertices are connected by undirected edges. 
\end{definition}
\begin{definition}[Maximal Independent set]
An independent set of a graph G is a subset S of vertices such that every pair of vertices are non-adjacent. An independent set is said to be maximal if no vertex can be added to S without violating independent set property.\end{definition}
\begin{definition}[Maximal Cliques]
A clique of a graph G is a subset S of mutually adjacent vertices. A clique is said to be maximal if no vertex can be added to S without violating clique property.\end{definition}
\begin{definition}[Chordal Graphs]
A graph is said to be chordal graph if every cycle of length at least four has a chord.
\end{definition}
\subsection{Conditional Independence and Markov Networks}
In this section we define conditional independence in probability(see \cite{Dawid} for more details), Markov network graphs and Markov networks. 
 
\begin{definition}[Conditional Independence] If $X, Y, Z $ are random variables with joint distribution P. Random variables X and Y are said to be conditionally independent given the random variable Z if and only if following holds.
\begin{align*}
	X  \bigCI Y \mid Z  &\iff P(X,Y \mid Z) = P(X \mid Z)P(Y \mid Z) \\
	& \iff P(X \mid Y,Z) = P(X \mid Z) 
\end{align*}
\end{definition}

\begin{definition}[Markov Network Graph]
A Markov network graph is an undirected graph G = ( V, E ) where $V= \{ X_1, X_2,..,X_n \}$ corresponds to random variables of a multivariate distribution. 
\end{definition}

\begin{definition}[Markov Networks] \label{definition:6}
A Markov network is a tuple $M = (G, \psi)$ where G is a Markov network graph, $\psi = \{ \psi_1, \psi_2, ..., \psi_m\}$ is a set of non negative functions for each maximal clique	$C_i \in G $ $\forall i = 1 \dots m $ and the joint pdf can be decomposed into factors as
\begin{align*}
	P(x) =\frac{1}{Z} \prod_{a \in C_m} \psi_a(x)
\end{align*}
where Z is a normalizing constant.
\end{definition}

\subsection{Graphical and Decomposable Log-linear Models}
In this section, we define a class of LLMs that can be represented by a graph.

\begin{definition}[Graphical Log-linear Models(GLLMs)]
A model is said to be $graphical$ if it contains all lower order terms which can be derived from variables contained in a higher-order term, then the model also contains the higher order interaction.
\end{definition}
For example, Let us consider a three-factor table. If a model includes all two factor interactions([12][23][13]) then it also contains the three factor interaction [123].  We usually represent a graphical model as a set of maximal cliques, which is [123] in this case.

\begin{definition} [Decomposable log-linear Models]
A model is decomposable if it is both graphical and chordal.
\end{definition}
 
\subsection{Model Checking and Model Comparison}
The goal of model selection is to choose a smallest graphical model from a class of graphical models under consideration that which best fits the data and has least number of edges(number of interaction terms). 
We use following test statistics for model testing and model comparison(see \cite{Christensen} and \cite{Whittaker} for further details).
\begin{itemize}
\item Pearson's $\chi^2$ Statistic:
It is defined as
\begin{align*}
	\chi^2 = \sum_i \frac{(O_i-E_i)^2}{E_i}
\end{align*}
where $O$ denotes observed cell count, $E$ as expected cell count.

\item The Deviance statistics:  In the generalized likelihood test, the test statistic is called deviance when we are comparing one model against saturated models otherwise for nested model its called the deviance difference. It is defined as follows.
\begin{align*}
	G^2 = -2 \sum_i O_i \log \frac{E_i}{O_i}
\end{align*}
Under null hypotheses the deviance is also distributes as $\chi^2$ with appropriate degrees of freedom.
\end{itemize}

\section{Mutual Conditional Independence in Markov networks}
\begin{theorem}
Let P be a positive distribution over V and let G be a Markov network graph over V, then elements of an independent set I of G are mutually conditionally independent given the rest $\{ V \setminus I \}$.
\end{theorem}

Here is an informal proof. Let us consider a simplest example a star graph as shown in figure(\ref{figure:1})
 Since the set $I = \{ B, C, D, E\}$ forms an independent set they belong to separate cliques  $[AB], \; [AC] \; [AD], [AE]$ respectively. From definition(\ref{definition:6}) the joint probability P factorizes as follows.
\[
	P = \psi_1(A, B) \; \psi_2(A, C) \;\psi_3(A, D) \psi_4(A, E)\\	
\]
The conditional probability $P(\{ B, C, D, E\} \mid A) $ can be expressed as 
\begin{align*}
	P(I \mid A =a) &= \psi_1(a, B) \; \psi_2(a, C) \; \psi_3(a, D) \; \psi_4(a, E)\\	
	&\implies \phi_1(B) \; \phi_2(C) \; \phi_3(D) \; \phi_4(E)
\end{align*}
Hence the factors $\{ B, C, D, E\}$ are mutually conditionally independent given the factor $A$ as given in figure (\ref{figure:2}). See \cite{Dan} for details on conditional independence for graphical models.

\begin{figure}[h!] 
\begin{minipage}{.4\textwidth}
\begin{center}
\[\begin{tikzpicture}
	\vertex (A) at (1,1) [label={30:$A$}] {};	
	\vertex (B) at (1,2) [label=above:$B$] {};  
	\vertex (C) at (0,1) [label=left:$C$] {};
	\vertex (D) at (2,1) [label=right:$D$] {};
	\vertex (E) at (1,0) [label=below:$E$] {};
	\path
		(A) edge (B)
		(A) edge (C)
		(A) edge (D)
		(A) edge (E)
	 ;   
\end{tikzpicture}\]
\caption{A Grphical Model} \label{figure:1}
\end{center}
\end{minipage}
\begin{minipage}{.6\textwidth}
\begin{center}
\[\begin{tikzpicture}
	\vertex (B) at (1,2) [label=above:$B$] {};  
	\vertex (C) at (0,1) [label=left:$C$] {};
	\vertex (D) at (2,1) [label=right:$D$] {};
	\vertex (E) at (1,0) [label=below:$E$] {};
\end{tikzpicture}\]
\caption{B,C,D,E are Mutually Separated by A} \label{figure:2}
\end{center}
\end{minipage}
\end{figure}

\section{Model Selection using Mutual Conditional Independence}
Finding an optimal graphical model from data sample is important mainly for prediction of future observables and describing the association among factors. 
As mentioned previously our goal is to find a simplest graphical model that fits the data. 
Now we present a new forward model selection procedure exploiting Mutual Conditional Independence Property(MCIP).

In this approach, we start with the null model (a complete independence model). We maintain two lists tempAMIS and AMIS. The tempAMIS contains a list of subset of factors to be tested for MCIP and the AMIS list contains MISs(for which the data supports MCIP). At each step we pick a set from tempAMIS, if its elements are mutually conditionally independent given the remaining factors then we remove it from tempAMIS and move it to the AMIS otherwise we find the most significant edge between them and add the required two factor and higher order terms into the present model. All sets in tempAMIS containing the end points of newly added edge gets split into two subsets. We repeat the process until there is no maximal independent set left to be tested for MCIP. The algorithm returns AMIS that determines the graph structure uniquely.
 
\begin{algorithm}[h!]
\SetAlgoLined
\textbf{Input:} A p-factor Contingency table, factors are labelled as $\{X_1,X_2, ... , X_p \}$\\
\textbf{Output:} AMIS:= All Maximal Independent Set\\
\caption{Forward Selection Algorithm}\label{fwd}
 \textbf{Initialize:}\\
 $AMIS = \phi$ \;
 $tempAMIS = \{\{ X_1,X_2, ... , X_p \} \}$\\
  \For { $ \textbf{ each } s \in tempMIS$}{
  \eIf{MCI relation holds amongst factors in s}{
  $tempMIS = tempMIS \; - \;s$ \;
  $AMIS = AMIS\; \cup \;s $ \; }
  { 
  find the most significant edge $(v_1,v_2) \in s$ \\
 	 \For  {$ \textbf{ each } t \in tempMIS$}
 	 {
 	   \If { $(v_1,v_2) \in t $} {
 			 $t_1 = t - v_1$ \;
			 $t_2 = t - v_2$ \;
 			 $tempMIS = \{tempMIS - t\} \cup t_1 \cup t_2$ \;
 			 }
 	 }
  } 
}
  \textbf{return} $AMIS$ 
\end{algorithm}

Let us consider a five-factor contingency table. Initial model assumption is the null model(all factor form an MIS), $tempAMIS = \{ \{1,2,3,4,5 \} \}$ and $AMIS = \{ \emptyset \}$. Let us suppose that the complete independence model does not fit and also assume that the edge $(1,2)$ is the most significant edge. We add this edge into the complete independence model. Since we had assumed that the set $\{1,2,3,4,5 \}$ is an MIS after adding the edge $(1,2)$ it is no longer an MIS, in fact now the assumption for the MISs are $tempAMIS = \{ \;\{\textbf{1},3,4,5 \}, \;  \; \{\textbf{2},3,4,5 \} \; \}$. Now let us suppose that the data supports the MCI condition for the set $\{1,3,4,5 \}$, we remove it from the tempAMIS and add it to the AMIS. At each step the procedure continues in this way until tempAMIS becomes empty.

Since it is more notational than conceptual, we begin the development with an example (\ref{Example:1}).

\begin{example} [Forward Model Selection for Rienis Dataset] \label{Example:1} We now illustrate the algorithm using real data Rienis dataset, for details on Reinis dataset see \cite{Reinis}. The Reinis data is shown in the table (\ref{table:1}). 
\begin{small}
\begin{longtable}[h!] {@{}ccccc|cccccc@{}} 
		\caption{Reinis data }\label{table:1}\\
		\toprule \centering
		&&&&& Smoke & \multicolumn{2}{c}{no} &  \multicolumn{2}{c}{yes}\\
		Family& Protein & Systol & Phys  & & Mental & no & yes &&no& yes \\
		\toprule \centering
		neg& $<3$ & $<140$ & no & & & 44 & 40 && 112 & 67\\
		&  &  & yes & & & 129 & 145 && 12 & 23 \\
		&  & $\geq140$ & no & &&  35 & 12 && 80 & 33\\
		&  &  & yes & & & 109 & 67 && 7 & 9 \\
		
		& $\geq3$ & $<140$ & no & & & 23 & 32 && 70 & 66 \\
		&  &  & yes & & & 50 & 80 && 7 & 13 \\
		&  & $\geq140$ & no & & & 24 & 25 && 73 & 57  \\
		&  &  & yes & & & 51 & 63 && 7 & 16 \\
		
		pos& $<3$ & $<140$ & no & &&  5 & 7 && 21 & 9 \\
		&  &  & yes & & & 9 & 17 && 1 & 4 \\
		&  & $\geq140$ & no & & & 4 & 3 && 11 & 8\\
		&  &  & yes & & &  14 & 17 && 5 & 2 \\
		
		& $\geq3$ & $<140$ & no & & & 7 & 3 && 14 & 14 \\
		&  &  & yes & & & 9 & 16 && 2 & 3 \\
		&  & $\geq140$ & no & &  & 4 &  0 && 13 & 11\\
		&  &  & yes & & &   5 & 14  && 4 &  4 \\
		\bottomrule \centering		
\end{longtable}
\end{small}
We begin by fitting the complete independence model, as given in figure(\ref{figure:3}). The vertices A,B,C,D,E,F correspond to the factors smoke,mental,phys,systol,protein,family respectively. The $G^2$ statistic for the model is $843.957$(df:57, p-value:0), hence we conclude that the data fails to support the complete independence model.
\begin{figure}
\centering
\begin{tikzpicture}
	\vertex (1) [label=above:A ]{};
  \vertex (2) [below  of=1] [label=below:B]{};
  \vertex (3) [ right of=1] [label=above:C] {};
  \vertex (4) [below  of=3] [label=below:D] {};
  \vertex (5) [right  of=3] [label=above:E]{};
  \vertex (6) [below  of=5] [label=below:F]{};
	\end{tikzpicture}
	\caption{The model of complete independence} \label{figure:3}
\end{figure}
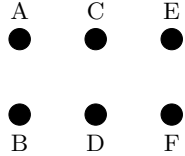

The data structures are initialized as follows.
\begin{align*}
	currModel &= [A] [B] [C] [D] [E] [F]\\
	tempAMIS & = \{ \; \{ A,B,C,D,E,F\} \; \}\\
	AMIS &= \{ \emptyset \}
\end{align*}

As mentioned before, at each step we add the most significant edge as long as the significance level is below a cutoff value(we use cutoff of $\alpha = 0.05$). As a first step we compare all the models with a single edge added to the model of complete independence. Table(\ref{table:2}) gives the model fit and table(\ref{table:3}) summarizes the test results.
\begin{small}
\begin{center}
\makebox[0pt][c]{\parbox{1.2\textwidth}{
\begin{minipage}[b]{0.6\hsize}\centering
\begin{longtable}[h!] {@{}clcc@{}} 
		\caption{Model Fitting } \label{table:2}\\
		\toprule \centering
		Ad. Edge & $\quad\quad$Model & d.f. & $G^2$   \\
		\toprule
		&   [A][B][C][D][E][F] & 57 & 843.9570 \\
		AB &[AB][C][D][E][F] & 56 & 834.2932 \\
		AC &[AC][B][D][E][F] & 56 & 816.4759 \\
		AD &[AD][B][C][E][F] & 56 & 832.9246   \\
		AE &[AE][B][C][D][F] & 56 & 826.5566   \\
		AF &[AF][B][C][D][E] & 56 & 842.8883   \\
		BC &[A][BC][D][E][F] & 56 & 157.9852   \\
		BD &[A][BD][C][E][F] & 56 & 843.4569   \\
		BE &[A][BE][C][D][F] & 56 & 826.0277   \\
		BF &[A][BF][C][D][E] & 56 & 839.2254   \\
		CD &[A][B][CD][E][F] & 56 & 843.8615   \\
		CE &[A][B][CE][D][F] & 56 & 827.2845   \\
		CF &[A][B][CF][D][E] & 56 & 843.7867   \\
		DE &[A][B][C][DE][F] & 56 & 831.1477   \\
		DF &[A][B][C][DF][E] & 56 & 842.8332   \\
		EF &[A][B][C][D][EF] & 56 & 840.9532  \\
		\bottomrule
\end{longtable}
\end{minipage}
\begin{minipage}[b]{0.32\hsize}\centering
\begin{longtable}[h!] {@{}ccrc@{}} 
		\caption{Model Comparision} \label{table:3}\\
		\toprule \centering
		Ad. Edge & d.f. & $G^2 \quad $  & p-value \\ \toprule
		AB & 1 & 9.6637 & 0.0018 \\
		AC & 1 & 27.4810 & 0.0000 \\
		AD &  1 & 11.0323 & 0.0008   \\
		AE  & 1 & 17.4003 & 0.0000   \\
		AF & 1 & 1.0686 & 0.3012   \\
		BC & 1 &  685.9717 & 0.0000   \\
		BD & 1 &  0.5000 & 0.4794   \\
		BE & 1 & 17.9292 & 0.0000   \\
		BF & 1 & 4.7315 &  0.0296 \\
		CD & 1 & 0.0954 & 0.7573   \\
		CE & 1 & 16.6724 & 0.0000   \\
		CF & 1 & 0.1702 & 0.6799   \\
		DE & 1 & 12.8092 & 0.0003   \\
		DF & 1 & 1.1237 &  0.2891   \\
		EF & 1 &  3.0037 & 0.0830  \\
		\bottomrule
\end{longtable}
\end{minipage}
}}
\end{center}
\end{small}
The model with edge $(B,C)$ has the largest difference in $G^2$ (or smallest p-value), we choose this models as the current model. Also the set containing the factors $(B,C)$ gets separated as follows.
\begin{align*}
	currModel &= [A] \textbf{[BC]} [D] [E] [F]\\
	tempAMIS & = \{ \;\{ A,\textbf{B},D,E,F\} , \; \{ A,\textbf{C},D,E,F \} \; \}\\
	AMIS &= \{ \emptyset \}
\end{align*}

As a next step, we consider the set$\{ A,B,D,E,F\}$.  
We perform the MCI test for the set and get $G^2$ statistic as $113.566$(df:52, p-value:0), which suggests that the data does not support the MCIP for the set. We compare the current model with all the models with an additional edge from the set $\{  A,B,D,E,F\}$. It is described in the table(\ref{table:4}) and table(\ref{table:5}). 

\begin{small}
\begin{center}
\makebox[0pt][c]{\parbox{1.2\textwidth}{
\begin{minipage}[b]{0.6\hsize}\centering
\begin{longtable}[h!] {@{}clcc@{}} 
		\caption{Model Fitting } \label{table:4}\\
		\toprule \centering
		Ad. Edge & $\quad\quad$Model & d.f. & $G^2$   \\ \toprule
		 &[A][BC][D][E][F] & 56 & 157.9852   \\
		AB & [AB][BC][D][E][F] & 55 & 148.3215 \\
		AD & [AD][BC][E][F] & 55 & 146.9529 \\
		AE & [AE][BC][D][F] & 55 & 140.5849   \\
		AF & [AF][BC][D][E] & 55 & 156.8614   \\
		BD & [A][BC][BD][E][F] & 55 & 157.4851   \\
		BE & [A][BC][BE][D][F] & 55 & 140.0559   \\
		BF & [A][BC][BF][D][E] & 55 & 153.2537   \\
		DE & [A][BC][DE][F] & 55 & 145.1760  \\
		DF & [A][BC][DF][E] & 55 & 156.9166   \\
		EF & [A][BC][D][EF] & 55 & 154.9815   \\
		\bottomrule
\end{longtable}
\end{minipage}
\begin{minipage}[b]{0.32\hsize}\centering 
\begin{longtable}[h!] {@{}ccrc@{}} 
		\caption{Model Comparision} \label{table:5}\\
		\toprule \centering
		Ad. Edge & d.f. & $G^2 \quad$  & p-value \\ \toprule
		AB & 1 & 9.6637 &0.0018 \\
		AD & 1 & 11.0324 & 0.0008 \\
		AE &  1 & 17.4003 & 0.0000   \\
		AF & 1 & 1.0686 & 0.3012   \\
		BD & 1 &  0.5000 & 0.4794   \\
		BE & 1 &   17.9292 & 0.0000   \\
		BF & 1 & 4.7315 &  0.02961\\
		DE & 1 & 12.8092 & 0.0003   \\
		DF & 1 & 1.1237 & 0.2891   \\
		EF & 1 & 3.0037 &  0.0830   \\
		\bottomrule
\end{longtable}
\end{minipage}
}}
\end{center}
\end{small}
The term $(B,E)$ is added to the current model. The data structures are updated as 
\begin{align*}
	currModel &= [A][BC]\textbf{[BE]} [D] [F]\\
	tempAMIS & = \{ \; \{ A,\textbf{B},D,F\},\;  \{ A,D,\textbf{E},F\} ,\;
	   \{ A,C,D,E,F \} \; \}\\
	AMIS &= \{ \emptyset \}
\end{align*}

Next we consider the set $\{ A,B,D,F\}$. The MCI test for the set gives the $G^2$ statistic as 67.5(df:44, p-value:0.013), we conclude that the data fails to support the MCI relation for the set.  
We consider an additional edge from the set $\{ A,B,D,F\}$, table(\ref{table:6}) gives the model fit and table(\ref{table:7}) summarizes the test results.

\begin{small}
\begin{center}
\makebox[0pt][c]{\parbox{1.2\textwidth}{
\begin{minipage}[b]{0.6\hsize}\centering
\begin{longtable}[h!] {@{}clcc@{}} 
		\caption{Model Fitting } \label{table:6}\\
		\toprule \centering
		Ad. Edge & $\quad\quad$Model & d.f. & $G^2$   \\ \toprule
		 & [A][BC][BE][D][F] & 55 & 140.0559   \\
		AB &[AB][BC][BE][D][F] & 54 &  130.3922 \\		
		AD &[AD][BC][BE][F] & 54 & 129.0236   \\
		AF &[AF][BC][BE][D] & 54 & 138.9873   \\
		BD &[A][BC][BD][BE][F] & 54 & 139.5558   \\
		BF &[A][BC][BE][BF][D] & 54 & 135.3244   \\
		DF &[A][BC][BE][DF] & 54 & 138.9321   \\
		\bottomrule
\end{longtable}
\end{minipage}
\begin{minipage}[b]{0.32\hsize}\centering
\begin{longtable}[h!] {@{}ccrc@{}} 
		\caption{Model Comparision} \label{table:7}\\
		\toprule \centering
		Ad. Edge & d.f. & $G^2 \quad$  & p-value \\ \toprule
		AB & 1 & 9.6637 &0.0018 \\
		AD &  1 & 11.0323 & 0.0008   \\
		AF & 1 & 1.0686 & 0.3012 \\
		BD & 1 &  0.5000 & 0.4795   \\
		BF & 1 &  4.7315 &  0.0296\\
		DF & 1 & 1.1237 &  0.2891   \\
		\bottomrule
\end{longtable}
\end{minipage}
}}
\end{center}
\end{small}
The term $(A,D)$ is added to the current model. The data structures get modified as 
\begin{align*}
	currModel &= \textbf{[AD]} [BC][BE] [F]\\
	tempAMIS & = \{ \;\{\textbf{A},B,F\} ,\; \{ B,\textbf{D},F\},\;  \{\textbf{A},E,F\}, \;	  \{\textbf{D},E,F\}, \;\{A,C,E,F\} \;, \{C,D,E,F\} \;\}  \\
	AMIS &= \{ \emptyset \}
\end{align*}

The $G^2$ statistics $38.915$(df:32, p-value:0.186) and  $39.271$(df:32, p-value:0.176) of the MCI tests for the sets $\{ A,B,F\}$ and $\{ B,D,F\}$ respectively indicates that the data supports the MCI relations for the sets. We remove them from tempAMIS and add to the AMIS.
\begin{align*}
	currModel &= [AD][BC][BE][F]\\
	tempAMIS & = \{ \; \{A,E,F\}, \;  \{D,E,F\}, \;\{A,C,E,F\} \;, \{C,D,E,F\} \;\}  \\
	AMIS &= \{ \;\{A,B,F\} ,\; \{ B,D,F\} \; \}
\end{align*}

Now we perform the MCI test for the set $\{ A,E,F\}$. The $G^2 \; 59.043$(df:32, p-value:0.002) statistic indicates that the data fails to support the MCI relation. We look for a significant edge in the set, the details are given in table(\ref{table:8}) and table(\ref{table:9}).
\begin{small}
\begin{center}
\makebox[0pt][c]{\parbox{1.2\textwidth}{
\begin{minipage}[b]{0.6\hsize}\centering
\begin{longtable}[h!] {@{}clcc@{}} 
		\caption{Model Fitting } \label{table:8}\\
		\toprule \centering
		Ad. Edge & $\quad\quad$Model & d.f. & $G^2$   \\ \toprule
		 &[AD][BC][BE][D][F] & 54 & 129.0236   \\
		AE &[AD][AE][BC][BE][D][F] & 53 & 111.6233 \\		
		AF &[AD][AF][BC][BE][D] & 53 & 127.9550   \\
		EF &[AD][BC][BE][D][EF] & 53 & 126.0199   \\
		\bottomrule
\end{longtable}
\end{minipage}
\begin{minipage}[b]{0.32\hsize}\centering
\begin{longtable}[h!] {@{}ccrc@{}} 
		\caption{Model Comparision} \label{table:9}\\
		\toprule \centering
		Ad. Edge & d.f. & $G^2 \quad $  & p-value \\ \toprule
		AE & 1 & 17.4003 & 0.0000 \\
		AF & 1 & 1.0686 & 0.3012   \\
		EF & 1 & 3.0037 &  0.0830 \\
		\bottomrule
\end{longtable}
\end{minipage}
}}
\end{center}
\end{small}
The term $(A,E)$ is added to the current model. The data structures get updated as follows(It must be noted that since $\{A,F\} \subset  \{A,B,F\}$ it is subsumed in it and its removed from tempAMIS). 
\begin{align*}
	currModel &= [AD]\textbf{[AE]}[BC][BE][F]\\
	tempAMIS & = \{ \; \{\textbf{E},F\}, \;  \;\{D,E,F\}, \;\{\textbf{A},C,F\} \;, \{C,\textbf{E},F\}, \;\{C,D,E,F\} \;\}  \\
	AMIS &= \{ \;\{A,B,F\} ,\; \{ B,D,F\} \; \}
\end{align*}
In the next step, we perform the MCI test for the set $\{ E,F\}$. We get $G^2$ statistics as $18.316$(df: 16, p-value:0.305), hence we conclude that the data supports MCI relation for the set. The tempAMIS and AMIS gets updated as follows.
\begin{align*}
	currModel &= [AD][AE][BC][BE][D][F]\\
	tempAMIS & = \{ \;\{D,E,F\}, \;\{A,C,F\} \;, \{C,E,F\}, \;\{C,D,E,F\} \;\}  \\
	AMIS &= \{ \;\{E,F\}, \; \{A,B,F\} ,\; \{ B,D,F\} \; \}
\end{align*}
Now we perform the MCI test for the set $\{ D,E,F\}$, the $G^2$ statistic is $49.428$(df:32, p-value:0.025), hence the data fails to support the MCI relation. We look for most significant edge between them. Test details are given in table(\ref{table:10}) and table(\ref{table:11}).
\begin{small}
\begin{center}
\makebox[0pt][c]{\parbox{1.2\textwidth}{
\begin{minipage}[b]{0.6\hsize}\centering
\begin{longtable}[h!] {@{}clcc@{}} 
		\caption{Model Fitting } \label{table:10}\\
		\toprule \centering
		Ad. Edge & $\quad\quad$Model & d.f. & $G^2$   \\ \toprule
		 & [AD][AE][BC][BE][F] & 53 &  111.6233   \\
		DE & [AD][AE][BC][BE][DE][F] & 52 & 93.3047 \\		
		DF & [AD][AE][BC][BE][DF] & 52 & 110.4995   \\
		EF & [AD][AE][BC][BE][EF] & 52 & 108.6195   \\
		\bottomrule
\end{longtable}
\end{minipage}
\begin{minipage}[b]{0.32\hsize}\centering
\begin{longtable}[h!] {@{}ccrc@{}} 
		\caption{Model Comparision} \label{table:11}\\
		\toprule \centering
		Ad. Edge & d.f. & $G^2 \quad $  & p-value \\ \toprule
		DE & 1 & 18.3185 & 0.0000  \\
		DF & 1 & 1.1237 & 0.2891   \\
		EF & 1 & 3.0037 &  0.0830 \\
		\bottomrule
\end{longtable}
\end{minipage}
}}
\end{center}
\end{small}

We select the model with an additional edge $(D,E)$. The set $\{D,E,F\}$ gets divided into the subsets $\{D,F\}$ and $\{E,F\}$. Since $\{D,F\} \subset \{ B,D,F\}$ and $\{E,F\}$ are already members of AMIS , hence the data supports MCI for them and therefore it is removed from the tempAMIS.

Similarly we proceed further and we find that the data fails to support the MCI relation for the set $\{A,C,F\}$. 
We compute following statistics for choosing the most significant edge between them.
\begin{small}
\begin{center}
\makebox[0pt][c]{\parbox{1.2\textwidth}{
\begin{minipage}[b]{0.6\hsize}\centering
\begin{longtable}[h!] {@{}clcc@{}} 
		\caption{Model Fitting } \label{table:12}\\
		\toprule \centering
		Ad. Edge & $\quad\quad$Model & d.f. & $G^2$   \\ \toprule
		 &[ADE][BC][BE][F] & 51 & 93.30472   \\
		AC &[AC][ADE][BC][BE][F] & 50 &  63.0128 \\		
		AF &[ADE][AF][BC][BE] & 50 & 92.2360   \\
		CF &[ADE][BC][BE][CF] & 50 & 93.1345   \\
		\bottomrule
\end{longtable}
\end{minipage}
\begin{minipage}[b]{0.32\hsize}\centering
\begin{longtable}[h!] {@{}cccc@{}} 
		\caption{Model Comparision} \label{table:13}\\
		\toprule \centering
		Ad. Edge & d.f. & $G^2$  & p-value \\ \toprule
		AC & 1 & 30.2918 & 0.0000  \\
		AF & 1 & 1.0686 & 0.3012   \\
		CF & 1 & 0.1702 &  0.6799 \\
		\bottomrule
\end{longtable}
\end{minipage}
}}
\end{center}
\end{small}

We choose the model with the edge $(A,C)$. The data structures get modified as 
\begin{align*}
	currModel &= \textbf{[AC]}[ADE][BC][BE][F]\\
	tempAMIS & = \{ \;\{C,F\} \;, \{C,E,F\}, \;\{C,D,E,F\} \;\}  \\
	AMIS &= \{ \;\{E,F\}, \; \{A,B,F\} ,\; \{ B,D,F\} \; \}
\end{align*}

On performing the MCI tests for the sets $(C,F) $ , $\{C,E,F\}$ and $\{C,D,F\}$ we get $G^2$ statistics as $22.153$(df:16, p-value:0.138), $42.534$(df:16, p-value:0.100) and $35.475$(df:16, p-value:0.307) respectively. It indicates that the data supports the MCI relations for them. The sets are removed from the tempAMIS and added to the AMIS. After removing redundant sets from the AMIS, finally we get the data structures as
\begin{align*}
	currModel &= [AC][ADE][BE][BC] [F]\\
	tempAMIS & =  \{ \emptyset \} \\
	AMIS &= \{ \; \{C,E,F\}, \;\{C,D,F\} \;  , \;\{A,B,F\} ,\; \{ B,D,F\} \; \}
\end{align*}

Since tempAMIS becomes empty, we stop with the model $[AC][ADE][BC][BE][F]$.
The algorithm returns the  $\{ \;\{ A,D,F\},\; \{C,E,F\}, \; \{B,E,F\}\; \}$ AMIS. A graph structure can be determined uniquely from the AMIS as given in figure(\ref{figure:4}).

\begin{figure}[h!]  
\begin{center}
	\begin{tikzpicture}
  \vertex (1) [label=above:A ]{};
  \vertex (2) [below  of=1] [label=below:D]{};
  \vertex (3) [ right of=1] [label=above:C] {};
  \vertex (4) [below  of=3] [label=below:E] {};
  \vertex (5) [right  of=3] [label=above:B]{};
  \vertex (6) [below  of=5] [label=below:F]{};
  \path[every node/.style={font=\sffamily\small}]
	(1) edge (2)
	(1) edge (4)			
	(2) edge (4)
    (1) edge (3)
    (5) edge  (3)
    (5) edge  (4);
	\end{tikzpicture}
	\caption{A Graphical Model for Reinis Data set} \label{figure:4} 
\end{center}		
\end{figure}
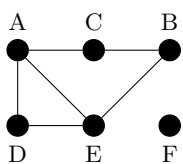
\end{example}
\section{Computational details}
All the experimental results in this paper were carried out using R 3.1.3. We implemented the new forward selection algorithm in R(Our implementation is available on request). We also used the existing packages gRim and MASS. All packages used are 
available at http://CRAN.R-project.org/.

\section{Conclusion and Future Scope}
In this paper, we have discussed mutual conditional independence property amongst the factors of a maximal independent set. We have presented an efficient forward model selection algorithm for the graphical log-linear models exploiting MCIP. 
Unlike traditional forward selection algorithm, our algorithm takes care of the evaluation of the joint effects since we perform mutual conditional independence check at every step for a group of factors therefore the important interactions can not be missed. \\
\\
We conclude with a couple of open problems as follows:\\
(i). The search space can be further reduced by considering only decomposable models(see \cite{Deshpande01efficientstepwise} for more details on stepwise selection in decomposable models).\\
(ii). The MCIP can be also used in backward elimination or in other stepwise selection procedures.

\end{document}